\documentclass[twoside,english]{elsarticle}
\usepackage[T1]{fontenc}
\usepackage[latin9]{inputenc}
\usepackage{graphicx}
\usepackage{esint}

\makeatletter

\providecommand{\tabularnewline}{\\}

\usepackage{lineno}

\makeatother

\usepackage{babel}
\begin{document}

\begin{frontmatter}{}

\title{Autonomous Inversion of In Situ Deformation Measurement Data for
$\mathrm{CO_{2}}$ Storage Decision Support}

\author[pnnl]{J. Burghardt\corref{cor1}}

\ead{jeffrey.burghardt@pnnl.gov}

\author[pnnl]{T. Bao}

\author[stanford]{K. Xu }

\author[UIUC]{A. Tartakovsky }

\author[stanford]{E. Darve }

\cortext[cor1]{Corresponding author}

\address[pnnl]{Pacific Northwest National Laboratory, Richland, Washington; USA
}

\address[UIUC]{University of Illinois Urbana-Champaign, Urbana-Champaign, Illinois;
USA}

\address[stanford]{Stanford University, Stanford, California; USA}
\begin{abstract}
Geologic carbon storage (GCS) is likely to play a key part of the
global effort to dramatically reduce $\mathrm{CO_{2}}$ emissions
and perhaps even reduce atmospheric $\mathrm{CO_{2}}$ concentrations
through carbon negative operations. A critical part of effort to commercialize
and widely deploy this technology is developing the capability to
rapidly assimilate real-time monitoring data into a form that will
enable site operators to make decisions to manage the safe and efficient
operations. Two of the risks associate with GCS are the risk of inducing
fractures in the sealing formations that can create leakage pathways
and the risk of inducing earthquakes of sufficient magnitude to cause
public concern, property damage, or safety risks. To properly manage
these risks the site operator needs to know the initial state of stress,
the change in stress induced by injection, and the relationship between
operational parameters such as injection rate and pressure and the
change in stress. Current methods of estimating the change in stress
require choosing the type of constitutive model and the model parameters
based on core, log, and geophysical data during the characterization
phase, with little feedback from operational observations to validate
or refine these choices. These characterization methods interrogate
the geologic formations using length scales, loading rates or magnitudes
that are quite different from those encountered by the actual storage
system. It is shown that errors in the assumed constitutive response,
even when informed by laboratory tests on core samples, are likely
to be common, large, and underestimate the magnitude of stress change
caused by injection. Recent advances in borehole-based strain instruments
and borehole and surface-based tilt and displacement instruments have
now enabled monitoring of the deformation of the storage system throughout
its operational lifespan. This data can enable validation and refinement
of the knowledge of the geomechanical properties and state of the
system, but brings with it a challenge to transform the raw data into
actionable knowledge. We demonstrate a method that uses automatic
differentiation and a finite-element based geomechanical model perform
a gradient-based deterministic inversion of geomechanical monitoring
data. This approach allows autonomous integration of the instrument
data without the need for time consuming manual interpretation and
selection of updated model parameters. Furthermore, only isotropic
linear elasticity is considered in this paper, the approach presented
is very flexible as to what type of geomechanical constitutive response
can be used. The approach is easily adaptable to nonlinear physics-based
constitutive models to account for common rock behaviors such as creep
and plasticity. The approach also enables training of machine learning-based
constitutive models by allowing back propagation of errors through
the finite element calculations. This enables strongly enforcing known
physics, such as conservation of momentum and continuity, while allowing
data-driven models to learn the truly unknown physics such as the
constitutive or petrophysical responses.
\end{abstract}

\end{frontmatter}{}

\section{Introduction}

Geologic carbon storage increases the pore fluid pressure and consequently
induces a deformation of the geologic strata. This deformation extends
well beyond the storage formation itself, and usually extends beyond
the upper and lower confining units. These deformations pose several
risks that a site operator must consider and manage. The two most
significant of these geomechanical risks are the risk of unintentionally
inducing shear or tensile fractures that compromise the integrity
of the sealing formations, and the risk of inducing earthquakes of
sufficient magnitude to cause public concern, property damage, or
safety risks. To properly manage these risks a site operator needs
to understand both the initial state of stress of the system, and
the changes in stress that are induced by storage operations. Traditional
approaches to characterizing the change in the system use physics-based
numerical geomechanical models. These are most commonly a finite-element
based model that is coupled in some manner to a reservoir model that
predicts the changes in fluid pressure caused by injection. 

A numerical geomechanical model most commonly solves the governing
equations for poroelastic continua. Some of these governing equations
are very well understood and well validated, such as conservation
of momentum, energy, and continuity. Unlike these well-known physical
laws that apply universally to all spatial locations, a geomechanical
analysis requires definition of a constitutive law that describes
the relationship between stress, pore pressure, and deformation in
each part of the domain being modeled. This relationship is not known
a priori but must be chosen for every spatial location at each site.
Characterizing this constitutive relationship, together with learning
the initial conditions, are the two key geomechanical challenges and
sources of uncertainty in accurately estimating changes is stress,
strain, and displacement that drive geomechanical risk. This work
will focus on a new method of learning the constitutive relationship
and leave the learning of the initial conditions out of the present
discussion.

First, we will address the question of what effect inaccuracies in
the constitutive relationship are likely to have on the practical
operation of a GCS site. When the pore pressure increases because
of $\mathrm{CO_{2}}$ injection this causes a decrease (i.e. becomes
less compressive) in the effective stress experiences by the rock
itself. This decrease in effective stress causes a volumetric expansion
of the pressurized rock. As one part of the subsurface expands adjacent
sections must also deform to accommodate this deformation. This will
cause some regions to compress and others to stretch. This deformation
and the associated change in stress is what is responsible for the
geomechanical risks of induced seismicity and the possibility of leakage
through induced or activated fractures. In general, the
larger the magnitude of these stress changes, the greater the geomechanical
risk. \citet{Burghardt2017b} used a tightly-coupled reservoir/geomechanical
model to compare two scenarios: one where the ratio of cap rock shear
modulus to reservoir shear modulus is 1.41, and one where this ratio
is 0.44. The results from that analysis showed that the more compliant
cap rock produces a change in stress that is approximately 50\% greater
than for the stiff cap rock case, which would indicate a much higher
geomechanical risk.

The example given above shows that a significant error in the shear
modulus estimate for the cap rock can lead to very large errors in
the geomechanical risk prediction. The next question to address is
whether such large errors in rock compliance are likely using existing
approaches. To minimize the time, cost, and complexity, the most common
characterization approaches assume a priori the simplest possible
constitutive relationship, usually isotropic linear elasticity. This
greatly simplifies the characterization problem, reducing the problem
to estimation of two elastic moduli (e.g. Young's modulus
and Poisson's ratio) that can be easily measured on
core samples and correlated with log and seismic responses. However,
by forcing the complex behavior of real subsurface systems into convenient
linear models, a limit is placed on the fidelity to which the model
can capture the reality of the subsurface processes.

\citet{Sone2014} conducted a series of tests to characterize the difference between
the deformation of US shale gas rocks during typical short-term laboratory
tests and two-week long tests. Their results showed that even after
two weeks of constant loading the samples were continuing to deform.
The time-dependent strain appeared to be following a logarithmic or
power-law type of response where the creep continues indefinitely
at a decreasing rate. Extrapolation using an assumed power law trend
showed that after one year the strain due to creep can reach 2.5 times
that that occurs due to the instantaneous elastic deformation. To
give a concrete example, if a sample of the Eagle Ford shale were
tested in the laboratory using ISRM suggested methods, a 10 MPa applied
stress would produce a strain of 0.4 millistrains, and result in a
Young's modulus of 25 GPa. If the same sample were
held under the same load for 1 year, using the creep compliance parameters
estimated by \citet{Sone2014}, the measured strain would be 1.4 millistrains
and the resulting Young's modulus would be 6.8 GPa.
A ten-year long test, which would be representative of the duration
of injection of a typical GCS site, would show an even lower Young's
modulus of 5.7 GPa. In other words, the short-term test that is almost
universally used today would overestimate the stiffness of the shale
sample by 365\% compared to how it would respond during the long-term
stress changes occurring during $\mathrm{CO_{2}}$injection. This
large error in the constitutive response would result in a very large
underestimation of the stress change and corresponding geomechanical
risk.

The results cited above show that errors in the constitutive relationship
current approaches are likely to be large (up to several hundred percent),
are not conservative (i.e. under-estimate risk), and to have important
consequences for the estimation and management of geomechanical risk.
A better estimation of the real constitutive response using current
state of the art methods, as employed by \citet{Sone2014} would result
in an increased characterization cost of several orders of magnitude
beyond what is common today since laboratory tests that take approximately
15 minutes today would be replaced with tests that take several hours
or even weeks. Additionally, there is reason to think that the estimate
given above are underestimates of the problem since core-scale tests
do not capture larger scale features such as natural fractures and
lithologic contacts that are additional sources of compliance and
creep.

Instead we propose using in situ geomechanical monitoring instrumentation
to update numerical geomechanical models throughout the lifespan of
an injection site. This is analogous to traditional history matching
approaches that use used to update the hydraulic properties of a reservoir
using injection and production data. \citet{Murdoch2020} recently
summarized recent advances in geomechanical instrumentation and demonstrated
that instruments are now available that can reliably resolve strains
as small as one nano-strain. Additionally, resent results from laboratory
\citep{Becker2019} and field tests \citep{Tan2021} show that fiber-optic
distributed acoustic sensing (DAS) measurements may be capable of
detecting very small magnitude strains, though extracting quantitative
quasi-static strains from DAS data is an ongoing research effort.
Even though some uncertainty remains about which strain sensing approach
will prove to be optimal considering trade offs of accuracy, cost,
and long-term reliability, it seems quite clear that data describing
the deformation of subsurface storage systems will soon be available
to operators. The challenge that remains is how to efficiently process
this data in such a way that a site operator can use to make operational
decisions.

\section{Method}

This section discusses the details of how a finite element method
is used to solve the inverse problem of estimating constitutive model
parameters using in situ geomechanical measurements. The method consists
of beginning with an initial estimate for the constitutive model parameters
and running standard finite element calculations to predict the monitoring
instrument responses. Errors between the simulated and predicted response
are quantified using a scalar loss function. The gradient of the loss
function with respect to the constitutive model parameters is then
computing by back propagation through the finite element model. Then,
using the loss function value and the its gradient with respect to
the inversion parameters, a quasi-Newton optimization algorithm can
be used to solve for the inversion parameters.

\subsection{Back propagation through finite element calculations}

A poromechanical finite element solver applies Newton-Raphson iteration
to find the nodal displacements, $u_{i}$, that satisfy static equilibrium
given body loads created by gravity and pore fluid pressure, constitutive
parameters, and boundary conditions. The finite element approximation
to the displacement field is constructed in such a way that the continuity
is automatically satisfied. The expression for the vector of nodal
residuals to the static equilibrium equation is

\begin{equation}
r(u_{i},\theta_{j})=\sum_{e=1}^{N_{e}}\intop_{\Omega_{e}}\left[B(x)\right]^{T}\cdot\sigma'(u_{i},\theta)dV_{e}-\intop_{\Omega_{e}}[B(x)]^{T}\cdot\alpha\delta P_{p}dV_{e}+\intop_{\Omega_{e}}\rho g[N(x)]dV_{e}\label{eq:discretized_momentum}
\end{equation}

where $N_{e}$ is the number of elements, $[B]$ is the strain-displacement
matrix for each element, $\sigma'$ is the Mandel effective stress
tensor, $\theta$ is a set of constitutive model parameters, $\alpha$
is the Biot coefficient, $P_{p}$ is the pore fluid pressure, $\rho$
is the mass density, $g$ is the acceleration due to gravity, and
$[N]$ is the shape function matrix for each element. Note that the
effective stress is a function of both the nodal displacements and
a set of constitutive parameters. As will be explained below, these
parameters may be traditional physics-based parameters such a Young's
modulus, creep compliance, yield strength, etc., or they may be weights
and biases in a deep learning model, or a combination of the two.
The Newton-Raphson iteration proceeds using the iteration formula

\begin{equation}
\left\{ u^{k+1}\right\} =\left\{ u^{k}\right\} -\left[J_{u}\right]^{-1}\left\{ r^{k}\right\} 
\end{equation}

Where $k$ is the iteration number, and $[J_{u}]$ is the Jacobian
matrix given by

\begin{equation}
J_{u}=\frac{\partial r(u_{i},\theta)}{\partial u_{j}}=\sum_{e=1}^{N_{e}}\intop_{\Omega_{e}}\left[B(x)\right]^{T}\cdot\left[\frac{\partial\Delta\sigma'(\theta)}{\partial\Delta\epsilon}\right]dV_{e}.\label{eq:FEM_Jacobian}
\end{equation}

If a set of measurements of the deformation field induced by injection,
$d^{i}$, are available, we can define a loss function, $L$, that
quantifies the degree to which the finite element solution agrees
with the measured response using an $L_{2}$ norm

\begin{equation}
L={\displaystyle \sum_{t=0}^{p}\sum_{i=0}^{m}}\left[\frac{d_{\mathrm{meas}}^{i}-d_{\mathrm{pred}}^{i}(t,u,\theta)}{S_{i}}\right]^{2}\label{eq:loss_function}
\end{equation}

where $p$ is the number of time steps and $m$ is the number of measurements
incorporated into the loss function, and $S_{i}$ is a scaling factor
for each measurement. The gradient of the loss function with respect
to the constitutive parameters can be expressed as

\begin{eqnarray}
\frac{\partial L}{\partial\theta} & = & \frac{\partial L}{\partial d_{\mathrm{pred}}}\frac{\partial d_{\mathrm{pred}}}{\partial u}\frac{\partial u}{\partial\theta}\nonumber \\
 & = & [L][M][G]\label{eq:dLdtheta}
\end{eqnarray}

The matrix $[L]$ is simple to calculate analytically from Eq. (\ref{eq:loss_function}).
The matrix $[M]$ is straightforward to calculate using standard finite
element shape functions for displacement measurements and strain-displacement
relationships for strain and tilt measurements. For displacement measurements
it is simply an identity matrix. Calculation of the derivatives for
the $[G]$ matrix is less straightforward since $u$ is an implicit
function of $\theta$ through the Newton-Raphson iteration procedure.
Following the approach presented by \citet{Xu2021a}, this obstacle
is overcome through application of the implicit function theorem,
which allows $[G]$ to be expressed as

\begin{eqnarray}
\frac{\partial u}{\partial\theta} & = & [G]=-\left[\frac{\partial r(u,\theta)}{\partial u}\right]^{-1}\left[\frac{\partial r(u,\theta)}{\partial\theta}\right]\nonumber \\
 & = & -[J_{u}]^{-1}\left[R\right]\label{eq:dudtheta}
\end{eqnarray}

Since $[J_{u}]$ is already computed as part of the Newton-Raphson
solution procedure, the only new calculation needed is the matrix
$[R]$, containing the derivatives of Eq. (\ref{eq:discretized_momentum})
with respect to the constitutive model parameters. In some cases,
such as for linear elasticity, this derivative can be computed analytically,
though the derivation and implementation is quite tedious and error
prone. For both purely physics-based models and for cases where the
constitutive model is replaced or augmented with a deep learning model,
the required derivatives can be computed using automatic differentiation
(AD). For the examples shown in this paper reverse-mode AD was implemented
using the PyTorch C++ library \cite{pytorch}. Unlike purely data-driven approaches, where 
AD is used to compute
gradients of a scalar quantity (the loss function) with respect to
a vector of parameters, in this physics-informed approach we require
computation of the Jacobian relating one vector (the nodal residuals)
with respect to a vector of parameters. This looses some of the computational
efficiency of reverse-mode AD (i.e. back propagation). Nevertheless,
reservoir mode AD is used for the analyses presented below, but further
research is recommended to determine the most efficient way to implement
AD to compute these derivatives.

Although the results shown below involve only single-process computations,
this approach lends itself to MPI-parallel computations, which are
expected to greatly speed up the process and be required for actual
CO2 storage site operations. For MPI-parallel finite element calculations
there is no need to use AD to back propagate across MPI processes
since the contributions of each process to $[R]$ can be computed
independently and summed during global assembly just as is done with
$[J_{u}]$. Although AD libraries capable of tracking derivatives across MPI 
processes now exist \cite{medipack}, it is possible that using such libraries
could improve the efficiency of the reverse-mode AD algorithm used here. This
is a recommended area for further research. Also, because the nodal displacements
are simply the sum
of the incremental change occurring during each time step, Eqs. (\ref{eq:dLdtheta})
and (\ref{eq:dudtheta}) can be applied to each time step and summed
along with the displacements to compute the total gradient after all
time steps to be used in training. Once the loss function and its gradients 
with respect to the constitutive model parameters have been calculated, a quasi-Newton optimization method can be used to solve the inverse problem for the constitutive
model parameters that best fit the data.

For the demonstration problem presented below the finite element calculations 
were implemented in a C++ code wrapped with a Python interface. The optimization
problem was then solved using the L-BFGS-B \citep{Byrd1995} optimization
algorithm provided by the SciPy optimization library. 

\section{Results}

\subsection{Description of case study problem}

A synthetic example problem was used to demonstrate the autonomous
data integration approach presented above. This example consists of
a simplified model of a GCS site. The model consists of a 20 m thick
reservoir at a depth of 1200 m. The sealing formation is a 200 m thick
unit. Shallower aquifers are represented by a single formation extending
from the ground surface to the top of the sealing formation. Underlying
the reservoir is a relatively stiff underlying formation extending
1000 m below the base of the reservoir. Table \ref{tab:formations}
lists the true elastic properties of each formation. The purpose of
this paper is to describe the model and demonstrate its use on a scenario
that is simplified from reality, but realistic in terms of the scale
and geometry of the problem and the number and magnitude of measurements.
For this reason we will limit the case study to homogeneous constitutive
model parameters within each formation, and limit the constitutive
model to isotropic linear elasticity.

As described in the introduction, one of the major motivations of this
work is the observation that real geomaterials seldom behave in an
isotropic linear elastic manner. This study should be seen as a step
toward the development of field instrumentation and numerical modeling
approaches that are able to characterize such complex rheology using 
in situ measurements. With that being said, there may be cases where
modeling the complex behavior of subsurface systems using a simplified
elastic model could be good enough to capture the coarse-grained response
of the system. Even a viscoplastic material response can be well
characterized with an elastic model if the loading rate is relatively
steady and of modest magnitude, such as a gradual increase of the pressure
in a GCS reservoir. Using an elastic model in this way amounts to fitting
the tangent stiffness of the system under the loading direction it is experiencing.
Such a model can be useful to extrapolate the behavior of the system
under continued loading of the same nature. Under this scenario, using
in situ instrumentation as proposed here would still offer a significant
advantage over use of core and log measurements because the loading
rates of the actual system are drastically different from those used
in core and log measurements. This approach would not generalize well
to changes in the nature of the system, such as depressurization.

\begin{table}
\begin{tabular}{|c|c|c|c|l|}
\hline 
Formation & Depth  & Young's  & Poisson's & Number of\tabularnewline
 & Range & Modulus & Ratio & Vertical Elements\tabularnewline
\hline 
Aquifer & 0-1000 m & 10 GPa & 0.3 & 6\tabularnewline
\hline 
Cap Rock & 1000-1200 m & 35 GPa & 0.28 & 8\tabularnewline
\hline 
Reservoir & 1200-1220 m & 20 GPa & 0.2 & 5\tabularnewline
\hline 
Basement & 1220-2220 m & 70 GPa & 0.23 & 6\tabularnewline
\hline 
\end{tabular}

\caption{\label{tab:formations}Four formations in model with their true elastic
properties}
\end{table}

Solving the pore pressure inversion problem is beyond the scope of
this paper, so this study takes the pore pressure field as deterministic
and well known. In practice, or course, this is not generally the
case and the determination of the pore pressure field, along with
the geomechanical state that is the focus of this paper, is one of
the primary challenges of $\mathrm{CO_{2}}$ storage site operation.
The pore pressure in the case study problem is prescribed to be initially
hydrostatic. The perturbation associated with injection modeled using
a simple linear function of the radius from the injection well and
is uniform across the vertical extent of the reservoir. The pressure
in all regions outside of the reservoir remain in a hydrostatic condition.
The center of the pressure plume is prescribed to increase linearly
with time such that the perturbation above hydrostatic pressure is
10 MPa at five years of injection. The edge of the pressure plume
is prescribed to be at 2.5 km from the injection well. 

\begin{figure}
\includegraphics{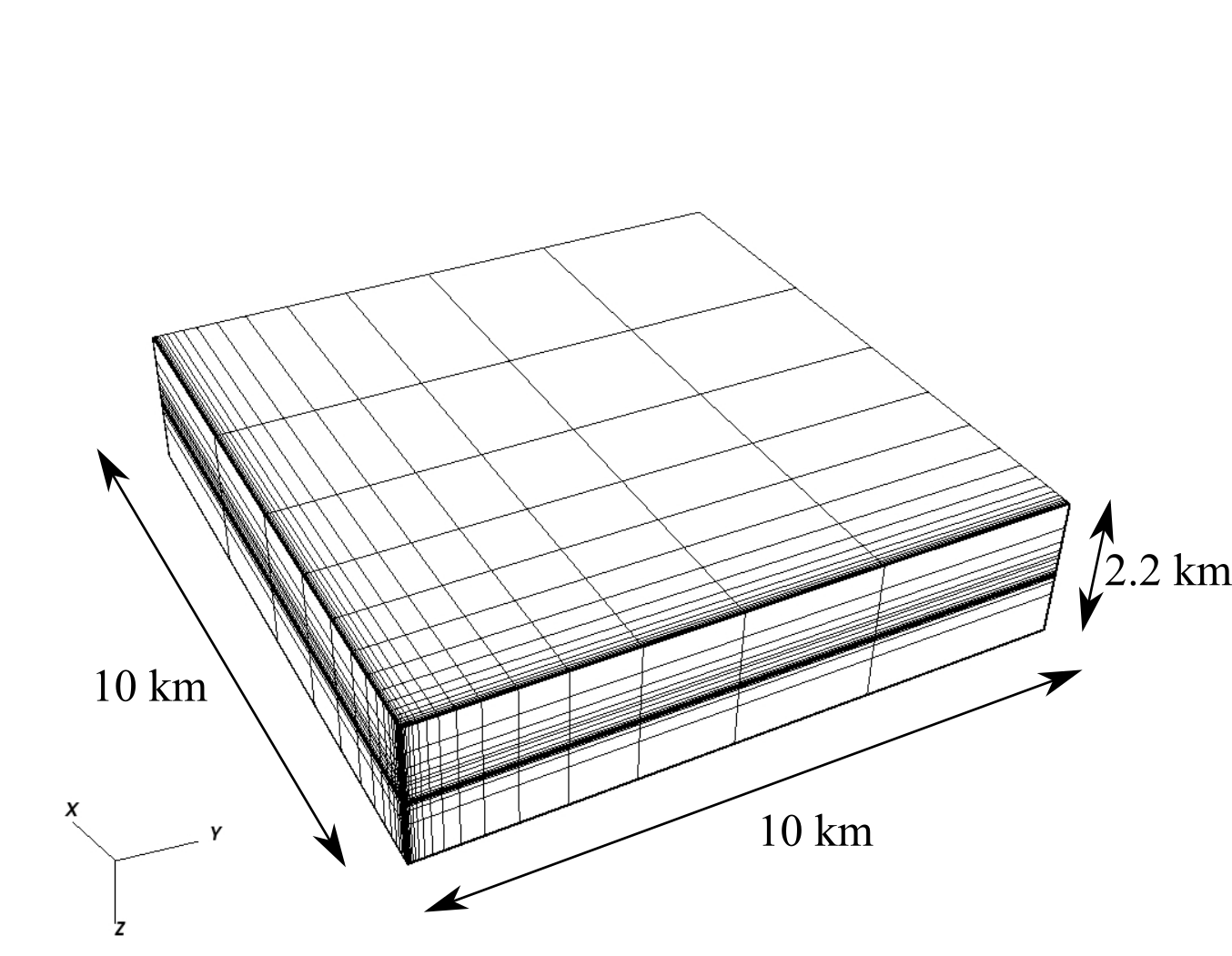}

\caption{\label{fig:mesh}The size of the computational domain and mesh used}

\end{figure}

Figure \ref{fig:mesh} shows the size of the computational domain
and mesh used for the calculations. The model represents a 1/4 symmetric
model of a 3D storage reservoir. Zero normal displacement boundary
conditions are specified on all except for the top surface, where
a zero traction boundary condition is applied. All formations are
assigned a uniform mass density of 2500 $\mathrm{kg/m^{3}}$ and gravitational
loading is applied in the z-direction, which is downward. The model
uses a structured hexahedral mesh with 20 elements in the lateral
dimensions and 25 in the vertical direction. The number of elements
used in the vertical direction for each formation is specified in
Table (\ref{tab:formations}).

To improve the conditioning of the optimization problem, in addition to scaling the
measurement contributions to the loss function as done in Eq. (\ref{eq:loss_function}),
the constitutive parameters are also scaled to be approximately of
order unity. Specifically, the Young's modulus was scaled by

\begin{equation}
\bar{E}=\frac{E}{E_{o}}.
\end{equation}
For all calculations presented below $E_{o}$ was chosen to be 100
GPa. Poisson's ratio was scaled by

\begin{equation}
\bar{\nu}=\frac{\nu}{0.5}
\end{equation}

Using the bounds available in the L-BFGS-B algorithm, $\bar{E}$ was
constrained to be greater than $10^{-3}$ and less than $1.0$, while
$\bar{\nu}$ was constrained to be greater than $0.0$ and less than
$0.9$. These bounds ensure positive definiteness of $[J_{u}]$ and
stability of the solutions to linear systems containing it.

\subsection{Geomechanical Measurements}

A set of geomechanical monitoring measurements that would be technically
feasible to implement were chosen to guide the autonomous data integration
and inversion. Table \ref{tab:measurements}, in the appendix, summarizes
the location, type, and maximum value of each of the 33 measurements
informing the inversion. One ground surface vertical displacement
measurement measurement is used in the inversion. This measurement
is located at the injection well location and hence center of the
pressure plume. Measurement such as this can be made using InSAR measurement.
While InSAR surveys can provide a rich 2D data set, we have chosen
to only use a single displacement measurement here because it is assumed
that the spatial distribution of the surface displacement field is
most sensitive to the pressure distribution, which we are taken as
a given in this study. The magnitude of the ground surface displacement,
however, is sensitive to the magnitude of the pressure perturbation
and the elastic property field. For this reason we have chosen to
use a single displacement point in the inversion since that will quantify
the magnitude of the surface displacement. The strain measurements
with the smallest magnitude used in the analysis are in the base formation
at the 3 km offset well, which have a magnitude of approximately 50
nano-strain ($n\epsilon)$. Resolving strains of this magnitude would
require strain meters grouted into the well bore, which have a resolution
of approximately 10 $n\epsilon$ \citep{Murdoch2020}. All but one
of the strains in the 1 km offset well have a magnitude of over 100
$n\epsilon$, which according to \citet{Murdoch2020} may be resolvable
with strain meters cementing in the casing annulus, and are definitely
within the rang resolvable by strain meters temporarily clamped in
the borehole. All strain measurement were scaled, according to Eq.
(\ref{eq:loss_function}), using a scale factor of 500 $n\epsilon$,
and the ground surface displacement was scaled by 5 mm.

\subsection{Inversion Results}

\begin{figure}
\includegraphics{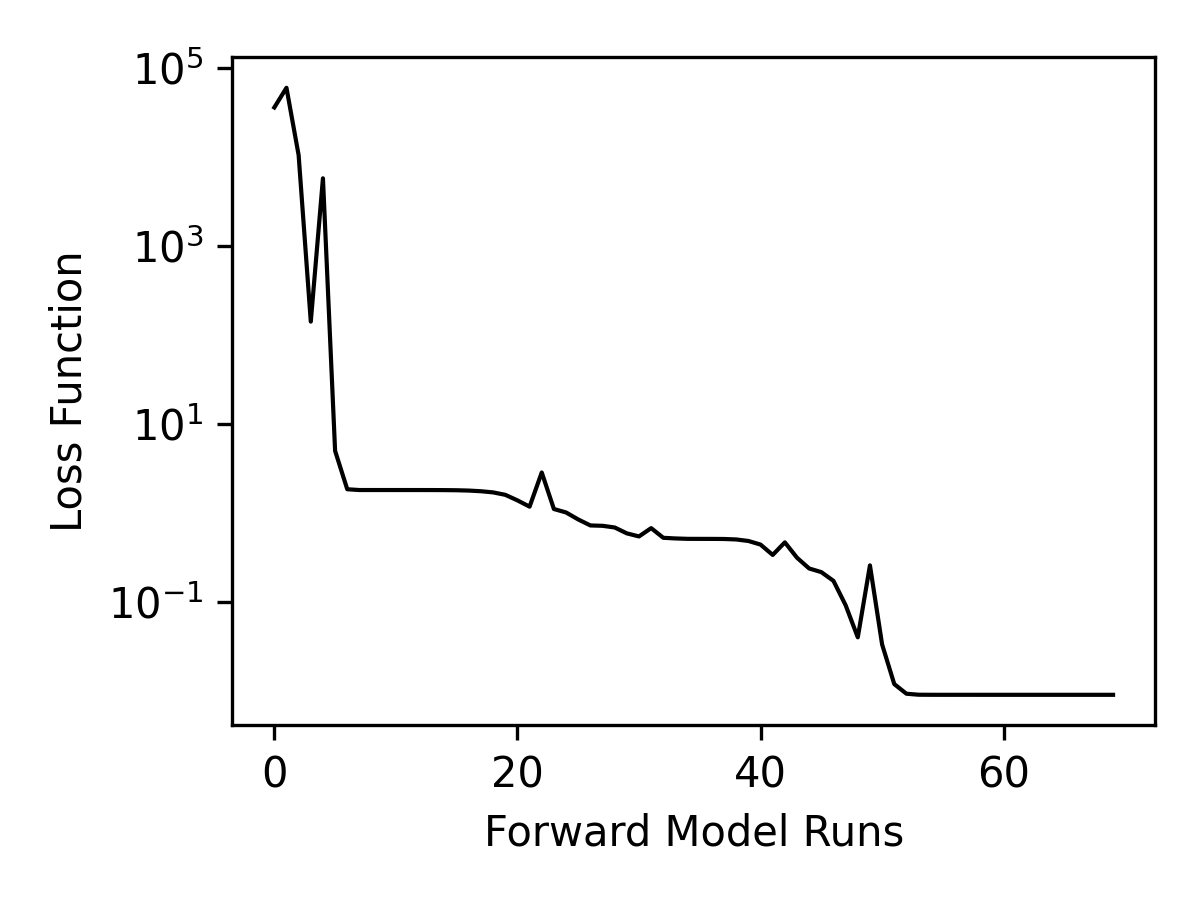}

\caption{\label{fig:loss_function_plot}Plot of the loss function value versus
number of forward model runs}

\end{figure}

The inversion was performed by initializing each parameter to the
middle of its range, $50$ MPa for the Young's modulus of each formation
and $0.25$ for each Poisson's ratio ($\bar{E}=\bar{\nu}=0.5$). As
a test of the algorithm initially only a single time step of 5 years
was used. Since the pressure perturbation in a self-similar over time,
and the material is linear elastic, adding additional time steps would
not add new information to the inversion. Figure \ref{fig:loss_function_plot}
shows a plot of the loss function value as a function of the number
of forward model runs. Note that the L-BFGS-B algorithm chooses a
search direction based upon the gradient from a limited number (10
in this case) past iterations, and then performs an inexact line search
in that direction, which required evaluating the forward model multiple
times for each iterations. Since the forward model runs are the most
computationally intensive part of the calculation all plots in this
section are made with respect to the number of forward model evaluations
rather than L-BFGS-B iterations.

\begin{figure}
\includegraphics{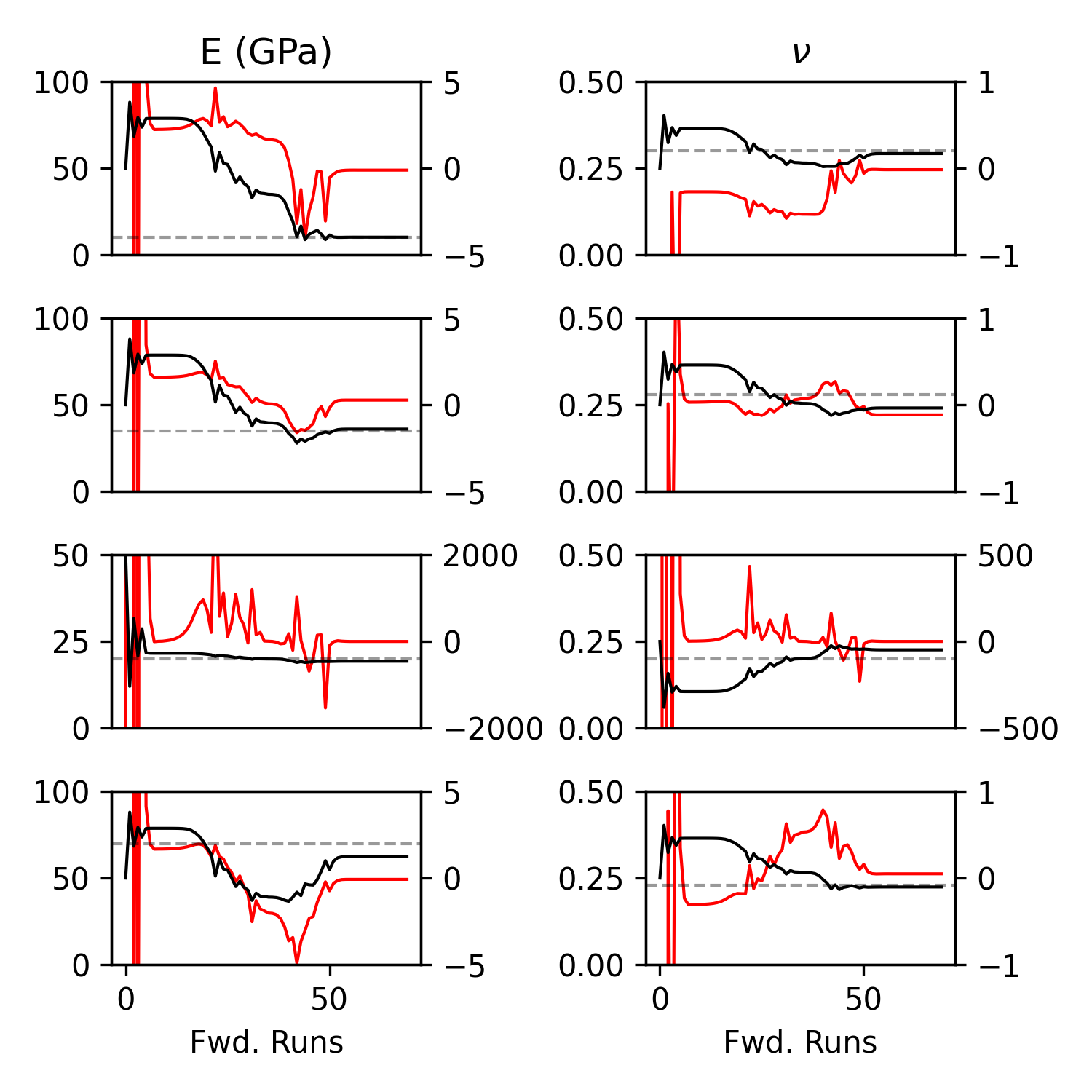}
\caption{\label{fig:inv_parameter_training}Plot of each of the nine inversion
parameters versus number of forward model runs. The black curves are
the parameter values, the dashed horizontal line is the true value,
and the red lines are the gradients of the loss function with respect
to the scaled version of each inversion parameter, which values plotted
on the right-hand axis.}
\end{figure}

Figure \ref{fig:inv_parameter_training} shows a plot of the value
of the Young's modulus and Poisson's ratio for the four formations
as a function of the number of forward model runs. As the figure shows,
each formation begins with the initial estimate of 50 GPa and 0.25.
The loss function is most sensitive, by several orders of magnitude,
to the reservoir elastic properties\textendash particularly the Young's
modulus. As a result of this the algorithm initially makes the largest
changes in this parameters and after approximately 25 forward model
evaluations this parameters has nearly converged to its true value.
The next most sensitive parameter, despite its relative distance from
the reservoir, is the Young's modulus of the shallow aquifer. This
is most likely do to the thickness of this formation (1 km), which
makes it have a large impact of the overall stiffness of the system.
If a more realistic model were constructed with multiple overlying
units it would be expected that the greater the distance from the
reservoir the lower the sensitivity, but as this example shows, the
thicker the formation the higher the sensitivity. More research needs
to be done to evaluate how much information is lost when formations
at a significant distance from the reservoir are lumped together into
a single effective continuum. It seems likely that this could, as
in this case, significantly simplify the problem perhaps without great
cost in accuracy.

\begin{figure}
    \centering
    \includegraphics{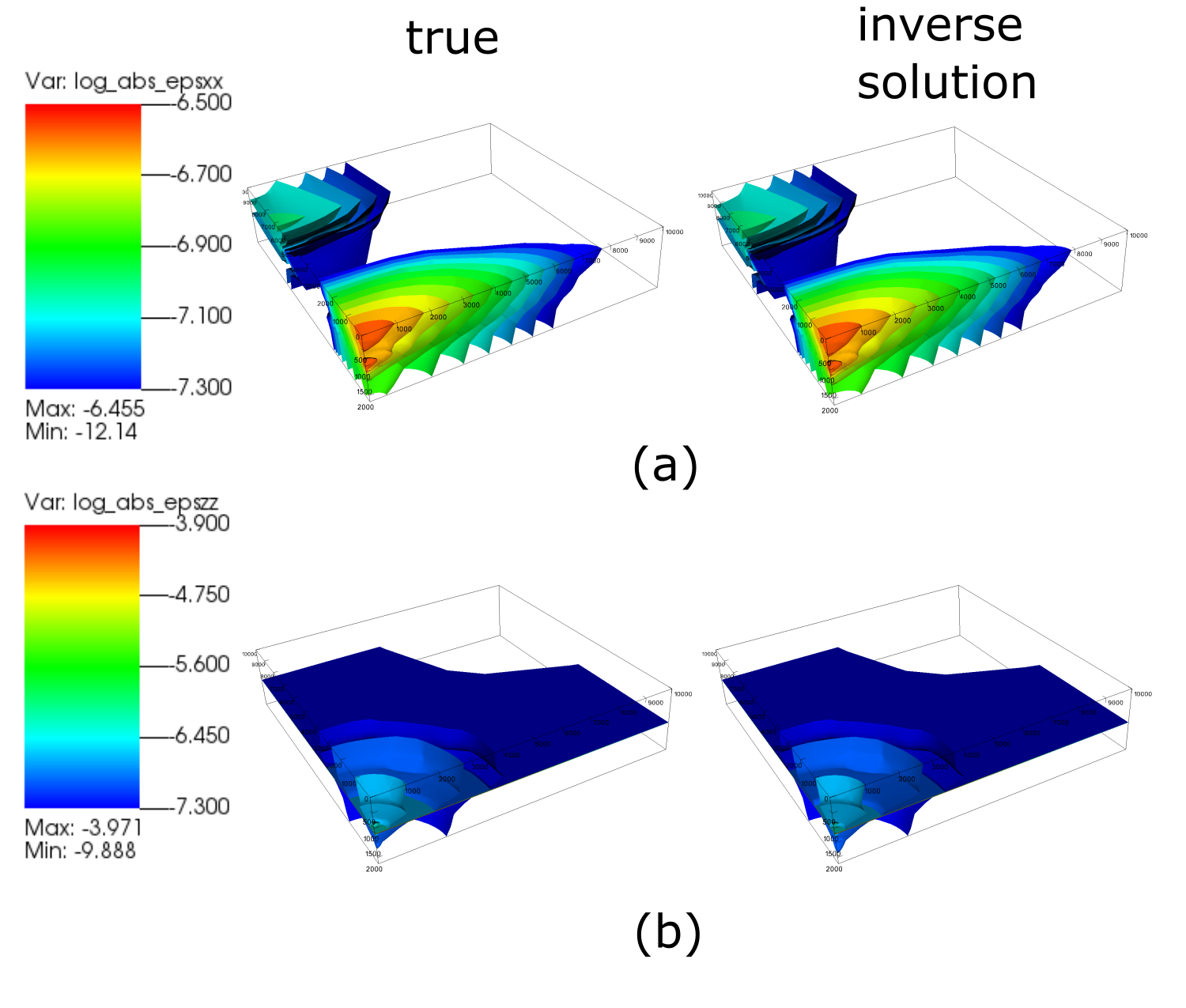}
    \caption{\label{fig:strain_solution} Comparison between the true strain solution between the true model and the inverse solution; (a) $\epsilon_{xx}$ component of strain; (b) $\epsilon_{zz}$ component of strain }
\end{figure}

\begin{figure}
    \centering
    \includegraphics{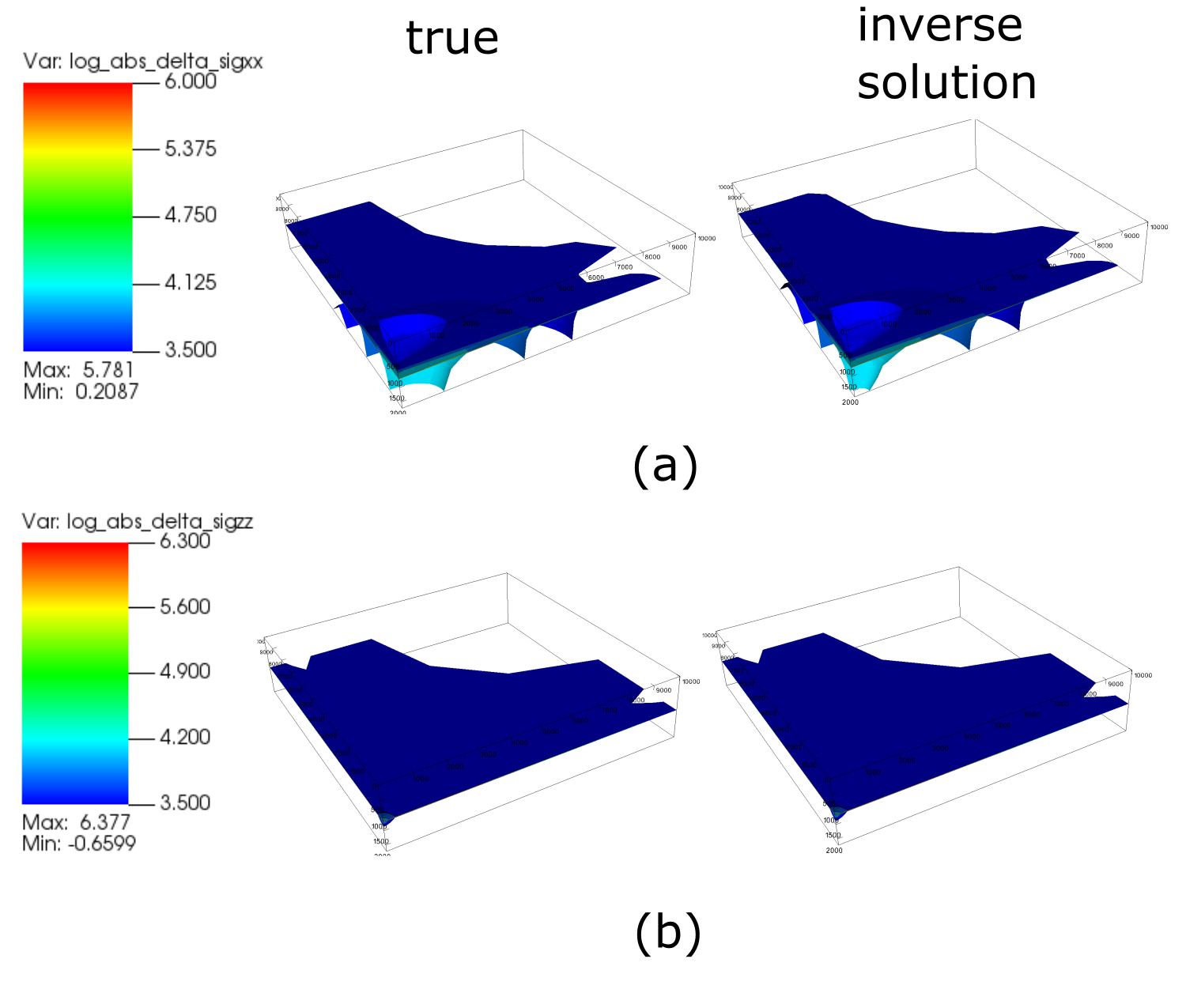}
    \caption{\label{fig:stress_solution} Comparison between the true stress solution between the true model and the inverse solution; (a) $\sigma_{xx}$ component of strain; (b) $\sigma_{zz}$ component of strain }
\end{figure}

Figures \ref{fig:strain_solution} and \ref{fig:stress_solution} show comparisons between the true model and the inverted model in terms of the change in strain and stress fields induced by injection, respectively. Because the change in stress and strain values vary by several orders of magnitude throughout the problem domain and have both positive and negative, the changes are plotted in terms of the base-ten logarithm of the absolute value of the change of each component. The normal components of stress and strain in the y-direction are not shown since the problem is axisymmetric. The off-diagonal components of stress and strain are not shown because they are only a few kPa since the principal stress and strain directions are close to the x-, y-, and z-axes. In all cases the strain components are accurate to within three significant figures and the stress components are accurate to within less than one MPa.

\section{Discussion}

One of the key barriers for widespread commercial deployment of GCS
is uncertainty about the ability to predict the hydromechanical response
of the storage system over long periods of time. This uncertainty
creates an elevation in perceived risk. This is particularly true
of geomechanical aspects of the response because, unlike hydraulic
aspects of the system, until recently there has been little or no
data to inform and update geomechanical model predictions from the
operational phase of a project. The primary diagnostics currently
available are ground surface displacement using InSAR, changes in
seismic attributes over time, and microseismic event locations and
source mechanisms. While InSAR measurements have proved very insightful
in a few cases {[}InSalah, Porotomo{]}, many injection sites have
failed to register any significant InSAR signals above background.
The surface expression of subsurface deformation is strongly influenced
by the depth of injection and the distribution of geomechanical properties
in the subsurface. Furthermore, many potential storage sites lie in
areas where agricultural operations such as harvesting and plowing
of fields, and seasonal aquifer draw down add noise that makes the
injection-related signal difficult to detect.

Imaging changes in stress from surface-based seismic data can be very
powerful in some cases, but it has several shortcomings. The first
significant limitation is temporal resolution. Because of the large
expense of repeated seismic surveys, time-lapsed seismic data is only
likely to be available with years between acquisitions. Second, is
the spatial resolution that is possible with surface-based seismic
surveys. Because seismic attenuation increases substantially with
frequency, deep seismic imaging relies on low frequency, long wavelength
components, which necessarily limits the spatial resolution. A third
major shortcoming of seismic imaging of stress is that even when significant
changes in seismic attributes are detected, it is difficult to uniquely
attribute the change to stress in a quantitative way. The primary
reason for this is that the change in seismic attributes with stress
varies strongly from formation to formation, and there is generally
no way to calibrate a seismic attribute-to-stress model since stress
changes generally cannot be measured directly. Laboratory tests on
core samples can be used to measure change in acoustic velocity with
stress, but it the differences in length scale and frequency that
are involved generally make this solution less than satisfactory.

Currently the most common way of providing feedback to geomechanical
models is using the locations, timing, and in some cases the source
mechanisms from microseismic events. This is most commonly done to
calibrate fracture propagation models in the petroleum industry. This
approach is highly non unique and often overlooks the complex relationship
between hydraulic fractures and microseismic events\textendash specifically
the fact that events can be triggered by a hydraulic fracture but
at a distance from the fracture itself. More rigorous approaches have
attempted to find relationships between in situ stress measurements
and focal mechanisms \citep{Cornet1989,Cornet1995}. These studies
have found that microseismic events often seem to be triggered at
local heterogeneities in the stress field, which biases the image
of the stress field if only this type of data is used. This is an
important point to consider, as it is both a weakness of using microseismic
events as a general stress field characterization tool, but it also
highlights that microseismic events are extremely useful for identifying
regions where the stress field is perturbed from larger-scale trends.
Such locations may be critical to identify and understand but by their
nature can be missed by stress measurement campaigns and coarse-grained
geomechanical models.

Despite the shortcomings of the three methods discussed above, all
of them have the potential to provide data that is informative for
determining the geomechanical properties and state of a subsurface
injection system under some circumstances. The strain measurements
that are the focus of this paper are also not without their shortcomings.
Rather than trying to decide which monitoring is best for informing
a geomechanical analysis it will be better to evaluate each method
at a particular site using a detailed feasibility analysis and developing
inversion algorithms that can use a wide variety of both redundant
and complementary data. 

\subsection{Notes on the deformation field and sensitivity to elastic properties}

The case study presented has a cap rock that is relatively stiff compared
to the reservoir. This is a likely scenario in many locations since
mature shale formations are often stiffer than high porosity sandstone
formations. However, there are locations where primary seals would
be composed of more compliant formations, such as immature shales
and reservoirs may be composed of more stiff formations such as moderate
porosity, well cemented sandstones or carbonate formations. Therefore,
the trends described here only apply to storage complexes similar
to the one considered here, where lateral strains due to expansion
of the reservoir are constrained by the relatively stiff cap rock
and underlying basement formation. 

Along the vertical center line of the pressure plume the vertical
strains are tensile inside the reservoir with a magnitude of approximately
250 micro-strain, and compressive outside the reservoir with a magnitude
of a few micro-strains. The lateral strains are all tensile, which
seems to be caused by the combined effect of lateral expansion of
the reservoir and expansion due to the Poisson effect associated with
compressive vertical strains outside of the reservoir. Because the
overlying formations are much more compliant than the underlying formations
there is also a bending deformation that occurs. Both the Poisson
effects and the bending effects seem to dominate in this case since
the lateral strains in the lower portion of the aquifer are larger
in magnitude than those in the lower portion of the cap rock. As mentioned
above, this is likely due to the constraint created by the relatively
stiff cap rock formation at this location. This same trend occurs
with the lateral strains at the 1 km offset monitoring well, where
strains outside of but within several hundred meters the reservoir
are on the order of several tenths of a micro-strain. The vertical
strains at 1 km offset are larger in the lower portion of the relatively
compliant aquifer formation than they are in the cap rock or base.
The relative magnitude of the Young's modulus in each formation is
a correlated more strongly than proximity to the pressurized reservoir.
Lateral strains follow a similar trend suggesting that the Poisson
effect is the stronger driver of lateral strains at this location
than is lateral stretching of the formations outside of the reservoir.

\subsection{Designing geomechanical monitoring approaches}

The geomechanical monitoring system used in the case study problem
presented here is only one of an infinite number of possible configurations,
and is likely far from optimal. In an early draft of this study the
inversion was attempted without strain monitoring in the injection
well, and convergence to the true state of the system was much slower
and less accurate. As described in the previous section, most of the
strains outside of the reservoir and along the periphery of the pressure
plume are caused by a combination of lateral stretching of the reservoir
and vertical compression caused by expansion of the reservoir. This
latter effect seems to dominate in the case considered here with relatively
stiff over- and underlying formations. The result is that the lateral
strains in the overlying formations are from a combination of the
Young's modulus and Poisson's ratio such that the inversion algorithm
may have a hard time converging on the correct value of the Young's
modulus, for example, because there are many pairs of Young's modulus
and Poisson's ratio that fit the data almost as well as the true values.
It was found that this effect was minimized by by including strain
values along the injection well where strains were larger in magnitude,
and along the lower portion of each formation, in as close of proximity
to the reservoir as possible, since strain magnitudes generally decay
with distance from the reservoir. Though strains measured from an
injection well are large and very informative, the injection of cold
$\mathrm{CO_{2}}$ in the same well bore is likely to create artifacts
in the strain measurements that may be challenging to remove. Further
investigation of these effects is warranted.

In the case study problem considered here, the true state of the system
was able to be determined from only the injection well and two monitoring
wells along a single radial direction. This was because the modeled
system was radially symmetric. Furthermore, the number of inversion
parameters was only nine because the subsurface system was lumped
into a relatively small number of formations, each assumed to be homogeneous.
This clearly represents an optimistic scenario. The approach presented
here needs to be tested with subsurface systems with a higher degree
of heterogeneity and complex rheology to evaluate how robust it is
and what sort of monitoring strategies are required to constrain a
more complex model.

Another aspect that needs to be considered in future evaluations of
this approach is the effect of instrument noise. The very sensitive
strain measurements envisioned in this study will also be sensitive
to earth tides and in some cases human caused changes such as variation
of fresh water aquifer levels. Such perturbations present both challenges
and opportunities. The challenge they present is that these effects
need to be either removed from the instrument responses or incorporated
into the numerical model. The opportunity presented by these factors
is that if they are incorporated into the numerical model they represent
another forcing function beyond the fluid injection that can aid in
resolving the system response more accurately\citep{Becker2019}.

Thus, in terms of computational time this approach is obviously not
faster than existing approaches. However, given the potential for
current approaches to significantly underestimate changes in stress
and therefore geomechanical risk, comparison of the computational
cost of the proposed approach with current deterministic forward modeling
approaches is no appropriate. A more fair comparison is with two other
approaches that rely on operational feedback to update the model over
time: stochastic inversion approaches \citep{Hanna2019} and manual
parameter updating \citep{Rutqvist2016}. A quantitative comparison
of these approaches using the same data set would be very helpful
to better determine the advantages and disadvantages of each. However,
based on published results is appears that the stochastic inversion
approach requires several orders of magnitude more forward model runs
than the deterministic inversion approach presented here. However,
the stochastic approach has two distinct advantages over the approach
presented here. First, for a Monte Carlo approach each model run is
independent so they can be performed in parallel rather than sequentially.
Second, the stochastic approach also can be used to estimate uncertainty
in the inversion rather than only providing the best fit solution.
A hybrid of these approaches may be best, where uncertainty can be
estimated but the inversion can also be informed with gradient information.
Manual parameter updating is time consuming in the sense that it requires
an expert practitioner to manually evaluate data, gain an intuition
for where the discrepancies in the model and observations are, and
then determine, generally through trial and error, how to adjust the
model parameters to fit the data. In this sense, even if the deterministic
or stochastic inversion approaches are more computational expensive,
they are likely to provide more timely results given that they do
require minimal human interact with observational data once the inversion
has been set up. These inversion approaches are also able to find
model parameters that describe the observational data very well but
that expert intuition may fail to find.

\section{Conclusion}

This lack of substantial operational feedback on geomechanical predictions
using existing approach adds to uncertainties from deficiencies in
current geomechanical characterization approaches stemming from differences
in loading rate, magnitude, and length scale between characterization
methods and actual loading conditions. The model presented here demonstrates
a proof of concept for leveraging recent developments in in situ deformation
monitoring instrumentation to provide critical feedback to improve
geomechanical modeling predictions over time. This has the potential
to significantly improve upon the accuracy, level and timeliness of
current approaches geomechanical modeling and risk estimation. While
the approach suggested here requires additional costs to a site operator
in the form of installation and operation of additional instrumentation,
the approach also has the possibility of drastically reducing geomechanical
uncertainties, which may justify their costs by allowing the system
to be operated at a higher capacity, allowing the post-injection site
care period to be shortened, and by alerting an operator to a hazardous
condition early when the problem is small and manageable. 

\section{Acknowledgements}
Funding for this research was provided as part of the Science-informed Machine learning to Accelerate Real Time decision making for Carbon Storage (SMART-CS) Initiative (edx.netl.doe.gov). Support for this initiative came from the U.S. DOE Office of Fossil Energy under DOE contract number DE-AC05- 76RL01830. PNNL is operated by Battelle for the U.S. DOE under Contract DE-AC06-76RLO1830.
This report was prepared as an account of work sponsored by an agency of the United States Government. Neither the United States Government nor any agency thereof, nor any of their employees, makes any warranty, express or implied, or assumes any legal liability or responsibility for the accuracy, completeness, or usefulness of any information, apparatus, product, or process disclosed, or represents that its use would not infringe privately owned rights. Reference herein to any specific commercial product, process, or service by trade name, trademark, manufacturer, or otherwise does not necessarily constitute or imply its endorsement, recommendation, or favoring by the United States Government or any agency thereof. The views and opinions of authors expressed herein do not necessarily state or reflect those of the United States Government or any agency thereof.

We thank Dr. Grant Brohmhal, Dr. Srikanta Mishra, Dr. Josh White, Dr. Catherine Yonkofski, Dr. Diana Bacon, Dr. Alex Hanna, and Casie Davidson for their advice and encouragement.

\bibliographystyle{plainnat}
\bibliography{CO2SiteGeomechanicalInversion}

\begin{thebibliography}{13}
\providecommand{\natexlab}[1]{#1}
\providecommand{\url}[1]{\texttt{#1}}
\expandafter\ifx\csname urlstyle\endcsname\relax
  \providecommand{\doi}[1]{doi: #1}\else
  \providecommand{\doi}{doi: \begingroup \urlstyle{rm}\Url}\fi

\bibitem[med()]{medipack}
Medipack.
\newblock https://github.com/SciCompKL/MeDiPack.

\bibitem[pyt()]{pytorch}
Pytorch c++ api.
\newblock https://pytorch.org/cppdocs/.

\bibitem[Becker and Coleman(2019)]{Becker2019}
Matthew~W. Becker and Thomas~I. Coleman.
\newblock Distributed acoustic sensing of strain at earth tide frequencies.
\newblock \emph{Sensors}, 19\penalty0 (9):\penalty0 1975, 2019.
\newblock ISSN 1424-8220.
\newblock \doi{10.3390/s19091975}.
\newblock URL \url{https://www.mdpi.com/1424-8220/19/9/1975/pdf}.

\bibitem[Burghardt(2017)]{Burghardt2017b}
J.~Burghardt.
\newblock Geomechanical risk analysis for geologic carbon sequestration.
\newblock In \emph{51st U.S. Rock Mechanics/Geomechanics Symposium}, volume
  ARMA-2017-0478, 2017.

\bibitem[Byrd et~al.(1995)Byrd, Lu, Nocedal, and Zhu]{Byrd1995}
Richard~H. Byrd, Peihuang Lu, Jorge Nocedal, and Ciyou Zhu.
\newblock A limited memory algorithm for bound constrained optimization.
\newblock \emph{SIAM Journal on Scientific Computing}, 16\penalty0
  (5):\penalty0 1190--1208, 1995.
\newblock ISSN 1064-8275.
\newblock \doi{10.1137/0916069}.
\newblock URL \url{http://epubs.siam.org/doi/10.1137/0916069?cookieSet=1}.

\bibitem[Cornet and Jianmin(1995)]{Cornet1995}
F.~H. Cornet and Yin Jianmin.
\newblock Analysis of induced seismicity for stress field determination and
  pore pressure mapping.
\newblock \emph{Pure and Applied Geophysics PAGEOPH}, 145\penalty0
  (3-4):\penalty0 677--700, 1995.
\newblock ISSN 0033-4553.
\newblock \doi{10.1007/bf00879595}.

\bibitem[Cornet and Julien(1989)]{Cornet1989}
F.H. Cornet and Ph. Julien.
\newblock Stress determination from hydraulic test data and focal mechanisms of
  induced seismicity.
\newblock \emph{International Journal of Rock Mechanics and Mining Sciences \&
  Geomechanics Abstracts}, 26\penalty0 (3-4):\penalty0 235--248, 1989.
\newblock ISSN 0148-9062.
\newblock \doi{10.1016/0148-9062(89)91973-6}.

\bibitem[Hanna(2019)]{Hanna2019}
Alexander~C. Hanna.
\newblock \emph{Stochastic Parameter Estimation of Poroelastic Processes Using
  Geomechanical Measurements}.
\newblock Thesis, 2019.

\bibitem[Murdoch et~al.(2020)Murdoch, Germanovich, Dewolf, Moysey, Hanna, Kim,
  and Duncan]{Murdoch2020}
Lawrence~C. Murdoch, Leonid~N. Germanovich, Scott~J. Dewolf, Stephen~M.J.
  Moysey, Alexander~C. Hanna, Sihyun Kim, and Roger~G. Duncan.
\newblock Feasibility of using in situ deformation to monitor co2 storage.
\newblock \emph{International Journal of Greenhouse Gas Control}, 93:\penalty0
  102853, 2020.
\newblock ISSN 1750-5836.
\newblock \doi{10.1016/j.ijggc.2019.102853}.

\bibitem[Rutqvist et~al.(2016)Rutqvist, Jeanne, Dobson, Garcia, Hartline,
  Hutchings, Singh, Vasco, and Walters]{Rutqvist2016}
Jonny Rutqvist, Pierre Jeanne, Patrick~F. Dobson, Julio Garcia, Craig Hartline,
  Lawrence Hutchings, Ankit Singh, Donald~W. Vasco, and Mark Walters.
\newblock The northwest geysers egs demonstration project, california part 2:
  Modeling and interpretation.
\newblock \emph{Geothermics}, 63:\penalty0 120--138, 2016.
\newblock ISSN 0375-6505.
\newblock \doi{10.1016/j.geothermics.2015.08.002}.
\newblock URL \url{https://doi.org/10.1016/j.geothermics.2015.08.002}.

\bibitem[Sone and Zoback(2014)]{Sone2014}
Hiroki Sone and Mark~D. Zoback.
\newblock Time-dependent deformation of shale gas reservoir rocks and its
  long-term effect on the in situ state of stress.
\newblock \emph{International Journal of Rock Mechanics and Mining Sciences},
  69:\penalty0 120--132, 2014.
\newblock ISSN 1365-1609.
\newblock \doi{http://dx.doi.org/10.1016/j.ijrmms.2014.04.002}.
\newblock URL
  \url{http://www.sciencedirect.com/science/article/pii/S1365160914000896}.

\bibitem[Tan et~al.(2021)Tan, Wang, Rijken, Hughes, Lim Chen~Ning, Zhang, and
  Fang]{Tan2021}
Yunhui Tan, Shugang Wang, Margaretha Rijken, Kelly Hughes, Ivan Lim Chen~Ning,
  Zhishuai Zhang, and Zijun Fang.
\newblock Geomechanical template for distributed acoustic sensing strain
  patterns during hydraulic fracturing.
\newblock \emph{SPE Journal}, 26:\penalty0 1--12, 2021.
\newblock \doi{10.2118/201627-PA}.

\bibitem[Xu et~al.(2021 (under revision))Xu, Tartakovsky, Burghardt, and
  Darve]{Xu2021a}
Kailai Xu, Alexandre Tartakovsky, Jeffrey Burghardt, and Eric Darve.
\newblock Learning viscoelasticity models from indirect data using deep neural
  networks.
\newblock \emph{Computer Methods in Applied Mechanics and Engineering}, 2021
  (under revision).
\newblock URL \url{https://arxiv.org/abs/2005.04384}.

\end{thebibliography}

\section*{Appendix}

\begin{table}
\begin{tabular}{|c|c|c|c|c|c|}
\hline 
Measurement & Type & Formation & Depth (m) & Offset (m) & \tabularnewline
\hline 
\hline 
1 & vertical displacement & Aquifer & 0 & 0 & 1.6 mm\tabularnewline
\hline 
2 & east strain & Aquifer & 980 & 0 & 312 $n\epsilon$\tabularnewline
\hline 
3 & vertical strain & Aquifer & 980 & 0 & -502 $n\epsilon$\tabularnewline
\hline 
4 & east strain & Cap Rock & 1190 & 0 & 290 $n\epsilon$\tabularnewline
\hline 
5 & vertical strain & Cap Rock & 1190 & 0 & -541 $n\epsilon$\tabularnewline
\hline 
6 & east strain & Reservoir & 1210 & 0 & 356 $n\epsilon$\tabularnewline
\hline 
7 & vertical strain & Reservoir & 1210 & 0 & 107 $\mu\epsilon$\tabularnewline
\hline 
8 & east strain & Base & 1230 & 0 & 241$n\epsilon$\tabularnewline
\hline 
9 & vertical strain & Base & 1230 & 0 & -322$n\epsilon$\tabularnewline
\hline 
10 & east strain & Aquifer & 980 & 1000 & 213 $n\epsilon$\tabularnewline
\hline 
11 & west strain & Aquifer & 980 & 1000 & 158$n\epsilon$\tabularnewline
\hline 
12 & vertical strain & Aquifer & 980 & 1000 & -204$n\epsilon$\tabularnewline
\hline 
13 & east strain & Cap Rock & 1190 & 1000 & 174$n\epsilon$\tabularnewline
\hline 
14 & west strain & Cap Rock & 1190 & 1000 & 145$n\epsilon$\tabularnewline
\hline 
15 & vertical strain & Cap Rock & 1190 & 1000 & -137$n\epsilon$\tabularnewline
\hline 
16 & east strain & Reservoir & 1210 & 1000 & 170$n\epsilon$\tabularnewline
\hline 
17 & west strain & Reservoir & 1210 & 1000 & 144$n\epsilon$\tabularnewline
\hline 
18 & vertical strain & Reservoir & 1210 & 1000 & 90$\mu\epsilon$\tabularnewline
\hline 
19 & east strain & Base & 1230 & 1000 & 166$n\epsilon$\tabularnewline
\hline 
20 & west strain & Base & 1230 & 1000 & 142$n\epsilon$\tabularnewline
\hline 
21 & vertical strain & Base & 1230 & 1000 & -98$n\epsilon$\tabularnewline
\hline 
22 & east strain & Aquifer & 980 & 3000 & 135$n\epsilon$\tabularnewline
\hline 
23 & west strain & Aquifer & 980 & 3000 & 711$n\epsilon$\tabularnewline
\hline 
24 & vertical strain & Aquifer & 980 & 3000 & -118$n\epsilon$\tabularnewline
\hline 
25 & east strain & Cap Rock & 1190 & 3000 & 118$n\epsilon$\tabularnewline
\hline 
26 & west strain & Cap Rock & 1190 & 3000 & 535$n\epsilon$\tabularnewline
\hline 
27 & vertical strain & Cap Rock & 1190 & 3000 & -88$n\epsilon$\tabularnewline
\hline 
28 & east strain & Reservoir & 1210 & 3000 & 117$n\epsilon$\tabularnewline
\hline 
29 & west strain & Reservoir & 1210 & 3000 & 52$n\epsilon$\tabularnewline
\hline 
30 & vertical strain & Reservoir & 1210 & 3000 & 56$\mu\epsilon$\tabularnewline
\hline 
31 & east strain & Base & 1230 & 3000 & 115$n\epsilon$\tabularnewline
\hline 
32 & west strain & Base & 1230 & 3000 & 53$n\epsilon$\tabularnewline
\hline 
33 & vertical strain & Base & 1230 & 3000 & -62$n\epsilon$\tabularnewline
\hline 
\end{tabular}

\caption{\label{tab:measurements}Summary of measurement informing inversion}
\end{table}

\end{document}